\documentclass{aa}
\usepackage{graphicx}
\usepackage{txfonts}
\usepackage{epstopdf}
\begin{document}

\title{Polarimetric studies  of carbon stars at high Galactic latitude}
\author{Aruna Goswami\inst{ } and  Drisya Karinkuzhi\inst{ }}
\institute{Indian Institute of Astrophysics, Koramangala II Block, 
Bangalore-560034, India\\
           \email{aruna@iiap.res.in}}
\date{Received 28 June 2012 / Accepted 21 October 2012}
\abstract
{ Very little is known about the polarimetric properties of
CH stars and carbon-enhanced metal-poor (CEMP) stars, although many of 
these objects have been studied in detail both photometrically and 
spectroscopically. }
{ We aim to derive polarimetric properties for a large sample of
CEMP stars and CH stars to fill this gap.}
{  Multiband polarimetric observations were conducted in the first run
for a sample of twenty-nine objects that include twenty-two CEMP 
 and CH stars and  seven polarization
standards. Estimates of  polarization  were obtained using standard 
procedures of  polarization calculation.}
{Five objects in our sample do not show any significant polarization
over the different colours of BVRI. For the rest of the objects the 
derived percentage polarization estimates are  $\le$ 1\%, 
and they are  found to exhibit  random behaviour with respect to the 
inverse of the effective wavelength of observations.
Polarization  also does not seem to have any correlation with the effective 
temperatures  of the  stars.}
{ Our polarimetric estimates indicate there are circumstellar 
envelopes around these stars that are spherically symmetric or 
envelopes with  little or no dust. In the plane of differential 
polarization,   defined as the difference between the maximum and 
the minimum polarizations within the $BVRI$-bands, versus their 
visual magnitude, the stars  appear to be   confined to  a narrow band. 
The implication  of this trend for understanding the nature of the 
circumstellar environment remains to be determined and requires 
detailed modelling. }

\keywords{stars: individual; 
stars: carbon stars;
stars: Late-type;
stars: polarization;
stars: variable stars;
stars:circumstellar matter
}

\authorrunning{Goswami \& Karinkuzhi}
\titlerunning{ Polarimetry of carbon stars  }
\maketitle

\section{Introduction}
Many late-type variables are known to exhibit intrinsic linear polarization
due to the scattering of starlight from circumstellar dust distribution in 
non-spherically symmetric envelopes. However, not much is known about 
the polarization properties of CH stars or  the carbon-enhanced 
metal-poor (CEMP) stars,  although 
both photometric and spectroscopic studies do exist for a large portion
of these objects. These stars  which are distributed  in a wide range of 
Galactic 
latitudes, are expected to have circumstellar envelopes that give rise to
linear polarization.  The first-ever estimates of V-band polarimetry 
for a sample of CEMP stars was reported by Goswami et al. (2010a). Within 
the small sample of stars they found two  distinct  groups, one with
p ${\le }$ 0.4\% and the other with p ${\ge }$ 1\%, and  separation into two 
groups could be  linked to the evolutionary properties of CEMP stars.
Time-dependent photometric and polarimetric studies of late-type Mira
variables have shown that  the decrease in the degree of polarization 
corresponds to an increase in brightness and vice versa (Dyck (1968)). 
Polarimetric studies  of normal carbon stars  (Kruszewski \& Gehrels (1968))
have also revealed a number of  important  features. Among these, the most
significant are  a)  a flat wavelength dependence of polarization 
in yellow to blue, b) time variability  in the wavelength dependence 
of polarization, c) time variability in both the degree of polarization 
and the position angle,  and d) close  correlation of polarization  
with the light variations. It would be useful to examine these properties 
in the case of  CEMP  and CH stars. CH stars are known to be radial 
velocity variables and members of binary systems (McClure (1984); 
McClure \& Woodsworth (1990)).
 It is known from the
literature that stars  that show intrinsic polarization are known  
variable stars;  however, the converse statement may not be true, and 
not all variable stars may show intrinsic polarization. 

Among the CEMP stars,  the largest fraction is  characterized  by 
abundance  patterns of neutron-capture elements (CEMP-s stars) that are 
compatible with the s-process in asymptotic giant branch (AGB) stars. 
Such abundance patterns  are also characteristics of those CH stars that are 
members of the Galactic halo (\cite{har85}). 
Abundances of carbon, nitrogen, and 
s-process elements are greatly enhanced in CH stars with 
[C+N/Fe] $\ge$ 1 (\cite{va92}). The enhancement of s-process 
elements are greater for the heavy s-process peak elements (Ba through Sm) 
than for the light s-process peak elements Sr, Y, and Zr. Orbital data, 
the abundances of C, N, O, and  carbon isotopic ratios, are consistent with 
a binary picture in which C and s-process elements are transferred onto 
the  CH star from the AGB star (\cite{va92}), where the 
$^{13}$C(${\alpha}$, n)$^{16}$O reaction acts   as  the source 
of neutrons for the s-process. Radial velocity variations imply that CEMP-s 
stars are also  members of binary systems (\cite{pr00}, \cite{luc05}),
and a production mechanism similar to the one applicable to CH stars 
is believed to hold good for CEMP-s stars (\cite{ma10} and references 
therein).

The identification of CH stars and CEMP-s stars as the same class  has
important implications for certain areas, such as  the AGB nucleosynthesis 
and Galactic enrichment of heavy elements due to low-mass metal-poor stars.  
Polarimetric properties could reflect upon the nature of the circumstellar 
environment and complement spectroscopic observations. In particular, 
polarization,  a characteristic property of stars evolving from the red 
giant stage to planetary nebula, can be used as an important indicator 
of stellar evolution. Despite the potential usefulness, almost no 
polarimetric data currently exists on CH and CEMP stars. It would be 
useful  to know how these objects compare in terms of their circumstellar 
environment using  polarization  properties. It is towards  this goal that
we have undertaken to study  a selected sample of these objects through 
polarimetric observations.

In the case of stars that are components of binary systems, as is believed 
to be the case for CH stars and CEMP-s stars, it is reasonable to expect 
consequent asymmetries or inhomogeneities in the circumstellar envelope,
 hence a measurable net polarization. We have carried  out the BVRI 
polarimetric studies  for a  sample of stars comprising  a few CH stars  
selected from the  CH star catalogue of \cite{ba96} and a few CEMP 
stars from \cite{ch01}. Unlike the normal carbon stars, with respect 
to the inverse of the  effective  wavelength of observation, the 
polarimetric  estimates of these objects are found to show  random 
behaviour. However, the differential polarization, defined as the 
difference between the observed maximum and minimum polarization 
within the $BVRI$-bands with respect to their visual magnitudes,  is 
found to exhibit   similar bahaviour by the CEMP stars and the CH 
stars. The implication of this observed  trend  in understanding  
the nature of the circumstellar environment  remains to be determined 
and requires detailed modelling.

In section 2, we present observations and data reduction procedures. 
In section 3, we  discuss the polarimetric calibration, and in 
section 4 we  present our results and discussions. Conclusions 
are drawn in section 5. 

{\footnotesize
\begin{table*}
{\bf Table 1: Photometric parameters of  programme stars}\\
\begin{tabular}{ c | c  | c | c | c | c | c | c | c | c  }
\hline
Star            &    RA(2000)   &    DEC(2000)     &   $l$     & $b$       &   V   &     J   &  H      &   K    & Dt. of obs\\
\hline
                &               &                        &          &           &       &         &         &        &          \\
BD $+$59d389$^*$  & 02 02 42.08  & $+$60 15 26.45     & 131.6653 &$-$1.4005  &9.07   & 6.535   &  6.203  & 6.023  & 02.02.2011  \\
HE 0310$+$0059    & 03 12 56.91  & $+$01 11 09.70       &178.9527  &$-$45.730   &11.56 & 9.871   &  9.296  & 9.196  & 01.02.2011 \\
HE 0518$-$2322    & 05 20 35.57  & $-$23 19 14.31      &225.6163  &$-$29.7391 & 12.78 & 11.151  &  10.672 & 10.568 & 02.02.2011  \\
HE 0519$-$2053    & 05 21 54.41  & $-$20 50 35.40   &223.0620  &$-$28.6177 & 13.70 & 11.965  &  11.399 & 11.296 & 01.02.2011  \\
                &              &                         &          &           &       &         &         &        & 02.02.2011 \\
HD 43384$^*$    & 06 16 58.70  & $+$23 44 27.27    &187.9943  &3.5289     & 6.29 & 5.187   &  5.093  & 4.975  & 01.02.2011  \\
                &              &                       &          &           &       &         &         &        & 29.03.2011 \\
HD 55496         & 07 12 11.37  & $-$22 59 00.61    &235.7353  &$-$5.9670  & 8.40  & 6.590   &  6.043  & 5.931  & 28.03.2011  \\
HD 65583$^*$     & 08 00 32.12   & $+$29 12 44.48      &192.1574  &26.950     & 6.94  &  5.539  &  5.170  & 5.095  & 28.03.2011 \\
HE 0915$-$0327     & 09 18 08.23  & $-$03 39 56.69      & 235.2581 &30.0879    &12.90  & 9.969   &  8.989  & 8.609  & 28.03.2011  \\
HE 0916$-$0037    & 09 18 47.65  & $-$00 50 34.56   &232.6277  &31.7927    & 12.80 & 11.226  &  10.782 & 10.658 & 01.02.2011 \\
HD 81192         & 09 24 45.33 & $+$19 47 11.86  &210.2033  &42.3555    & 6.53 & 4.846   &  4.282  & 4.119  & 28.03.2011  \\
HD 90508$^*$    & 10 28 03.88 & $+$48 47 05.64    &165.1026  &54.9211    & 6.44  & 5.20    &  4.890   & 4.870   & 01.02.2011  \\
HE 1030$-$1518     & 10 33 10.06  & $-$15 33 50.26     &260.6244  &35.7077    & 12.06 & 10.375  &  9.801  & 9.657  & 29.03.2011  \\
HD 92545         & 10 40 57.70 & $-$12 11 44.21  &259.8507  &39.5218    & 8.56  & 7.548   &  7.347  & 7.282  & 28.03.2011  \\
HD 94851$^*$    & 10 56 44.25 & $-$20 39 52.62   &269.8199  &34.7282    & 9.27  & 8.869   &  8.816  & 8.771  & 02.02.2011  \\
                &               &                      &          &           &       &         &         &        & 29.03.2011 \\
HE 1056$-$1855    & 10 59 12.25  & $-$19 11 06.99     &269.4844  &36.2935    & 12.49 & 10.784  &  10.249 & 10.090 & 02.02.2011  \\
HD 100764       & 11 35 42.74 & $-$14 35 36.66     &276.8590  &44.4108    & 8.73  & 7.048   &  6.600  & 6.153  & 02.02.2011  \\
LEE 107         & 11 55 39.74   & $+$12 34 48.0      &258.2830  &70.4485    & 11.3  & 7.709   &  6.853  & 6.560  & 02.02.2011  \\
HD 107574        & 12 21 51.86 & $-$18 24 00.15  &293.1743  &43.9107    & 8.55  & 7.660   &  7.460  & 7.415  & 28.03.2011  \\
GD 319$^*$      & 12 50 05.00   & $+$55 06 00.00    &123.3446  &62.0271    &12.32  &11.548 & 11.087  & 11.056  & 02.02.2011  \\
HD 111721        & 12 51 25.19 & $-$13 29 28.17   &302.9257  &49.3806    & 7.97  & 6.347   &  5.898  & 5.786  & 28.03.2011  \\
HD 112869        & 12 59 22.64 & $+$37 49 03.69   &114.5490  &79.1809    &9.06   & 6.265   &  5.543  & 5.241  & 29.03.2011  \\
HE 1345$-$2616     & 13 48 02.08   & $-$26 31 12.40    &318.3520  &34.6512    & --    & 9.855   & 9.276   &  9.099 & 29.03.2011  \\
HD 121447        & 13 55 46.96 & $-$18 14 56.49  &323.6762  &42.0161    & 7.81  & 5.397  & 4.442  & 4.282  & 29.03.2011  \\
HD 125079        & 14 17 20.71 & $-$04 15 57.81    &339.6595  &52.3705    & 8.67  & 7.136   &  6.759  & 6.610  & 29.03.2011  \\
HD 126681        & 14 27 24.91 & $-$18 24 40.44    &332.6355  &38.8618    & 9.32  & 8.044   &  7.709  & 7.631  & 29.03.2011  \\
HE 1428$-$1950    & 14 30 59.39  & $-$20 03 41.90     &332.6046  &37.0084    & 11.86 & 9.988   &  9.469  & 9.318  & 01.02.2011  \\
HE 1429$-$1411     & 14 32 40.57  & $-$14 25 05.61    &336.6322  &41.7343    & 11.10 & 7.622   &  6.721  & 6.346  & 28.03.2011 \\
HD 160529$^*$    & 17 41 59.02 & $-$33 30 13.71   &355.7023  &$-$1.7325    & 6.77  & 3.547   &  3.056  & 2.790  & 28.03.2011  \\
                &               &                 &          &           &       &         &         &        & 29.03.2011 \\ 
HD 164922        & 18 02 30.8624 & $+$26 18 46.80   &52.2980   &21.7853    & 6.99  & 5.550   &  5.20   & 5.110  & 29.03.2011  \\
                &               &                 &          &           &       &         &         &        &     \\ 
\hline
\end{tabular}

 $^*$ indicates polarization standard star
\end{table*}
}

\section{Observations and data reduction}
The sample of stars observed  are listed in Table 1 along with several 
physical  and photometric parameters for each star. BVRI polarimetric 
observations  were carried out using the polarimeter attached to the 
2-m telescope at IGO, Girawali of IUCAA, Pune during February to  March
2011. In addition to the programme stars, a few standard polarization  
and  zero-polarization stars were observed for polarimetric calibrations. 
The polarimeter has  a rotating half wave plate (HWP) and Wollaston 
prism through which light passes before forming a pair of images of 
an object on the CCD (\cite{sen94}; \cite{ra98}). The HWP can rotate in
several discrete steps, such that its fast axis makes angles ($\alpha$) 
with some reference direction (generally celestial north-south). The 
light  transmitting out  of the Wollaston prism forms two images of the  
source on the CCD, with the ordinary and extraordinary set of rays. 
The observations for each object were taken at four different positions 
of the HWP:  0$^{0}$, 22.5$^{0}$, 45$^{0}$, and 67.5$^{0}$. Data reduction 
was carried out using various  tasks in IRAF. The task PHOT in APPHOT was 
used to measure the stellar flux. 

\section{Polarimetric calibration}
Three polarized standard  and four   unpolarized standard stars  
were observed during the observing run for polarimetric calibration.  
The  estimated  p\%   for the polarization standards are  listed in 
Table 2 along with their literature values whenever available. We find 
 close agreement between our  estimates and the literature values. 
For the standard polarized stars,  the position angles ($\theta$) are 
listed in column 5 of Table 2. For stars with null polarization the 
position angles are not defined, hence position angles are only presented 
 for stars that give non-zero polarization. 

 The errors in the polarization measurements are primarily by dominated
 photon noise. For low values of polarization,  the number of 
photo-electrons corresponding to ordinary and extraordinary images 
are approximately equal. The error in the polarization measurements 
are given by 
 ($\frac{\sqrt{N+sN_b}}{2N} \times 100\%$) where
$s$ = $\pi$r$^2$,  `r' is the radius of the aperture used to find 
the total flux of the star image (in pixels), N is the total source 
counts, and N$_{b}$  the average background sky count per pixel. The 
estimates of the percentage polarization (p\%),  along with the corresponding
error estimates,  are listed  in Table 3. For the majority of the objects 
the measured errors are  ${\le}$ 0.1\%. 

The zero-polarization, standard stars were observed to check for any 
possible instrumental error,  which proved to be $\sim$  0.1\%. 
As noted from Table 2,  the amount of polarization for the 
standard unpolarized  stars  HD~90508, HD~94851, HD~65583, and  GD~319 
are non-zero. It  is reasonable to expect the amount of polarization  
determined from a single measurement of a standard star to lie within 
the limits 0.0 ${\le}$ p\% ${\le}$ 0.21 (Dyck \& Jennings 1971).
Following this criterion  the objects with p  ${\le}$ 0.2\% are considered
to be  unpolarized at the epoch of observation.

{\footnotesize
\begin{table*}
{\bf Table 2:  Polarimetry of polarization standard stars }\\
\begin{tabular}{  c | c | c | c   c | c  c | c }
\hline
Star names     &  HJD           &Filter &     p          & $\theta$     &  p           & $\theta$ & Reference \\
               &                &       &     ($\%$)          & (degree)        &  ($\%$)         &  (degree)       &           \\
(1)            &  (2)           &  (3)  &     (4)           &     (5)        &  (6)          &   (7)          & (8)     \\
\hline
                &                &       &                   &                &               &                &     \\
HD~43384       & 2455594.31754  & B     & 2.92 $\pm$ 0.05   & 167.4        &                 &                &      \\
               &                & V     & 2.80 $\pm$ 0.06   & 172.6        & 3.0             &  170.0         & \cite{se75}    \\
               &                & R     & 2.91 $\pm$ 0.08   & 169.8        &                 &                &     \\
               &                & I     & 2.34 $\pm$ 0.07   & 169.3        &                 &                &     \\
HD~43384       & 2455650.12855  & B     & 3.10 $\pm$ 0.07   & 164.3        &                 &                &     \\
               &                & V     & 2.78 $\pm$ 0.08   & 172.2        &                 &                &     \\
               &                & R     & 2.81 $\pm$ 0.08   & 170.9        &                 &                &     \\
               &                & I     & 2.53 $\pm$ 0.11   & 171.9        &                 &                &     \\
\hline
HD~90508       & 2455594.33936  & B     &  0.14 $\pm$ 0.06  &              &                   &                &     \\
               &                & V     &  0.03 $\pm$ 0.08  &              &                   &                &     \\
               &                & R     &  0.08 $\pm$ 0.08  &              &                   &                &     \\
               &                & I     &  0.11 $\pm$ 0.07  &              &                   &                &     \\
\hline
GD~319         & 2455594.51665  & V     & 0.06 $\pm$ 0.04   &            & 0.04 $\pm$ 0.04 &                & \cite{tu90} \\
\hline
BD+59d389      & 2455595.08535  & B     & 6.79 $\pm$ 0.12   & 92.6    &                   &                &       \\
               &                & V     & 6.52 $\pm$ 0.05   & 101.7   & 6.67 $\pm$ 0.03  & 98.2           &  \cite{tu90} \\
               &                & R     & 6.40 $\pm$ 0.05   & 99.93    &                   &                &    \\
               &                & I     & 5.69 $\pm$ 0.05   & 99.02    &                   &                &       \\
\hline
HD~94851       & 2455595.28591  & B     & 0.09 $\pm$ 0.08  &          &                   &                &   \\ 
               &                & V     & 0.09 $\pm$ 0.10  &          & 0.06  $\pm$ 0.02 &                & \cite{tu90}\\ 
               &                & R     & 0.21 $\pm$ 0.12  &          &                   &                &         \\
               &                & I     & 0.19 $\pm$ 0.18  &          &                   &                &      \\
HD~94851       & 2455650.19658  & B     &  0.10 $\pm$ 0.05  &        &                   &                &      \\
               &                & V     &  0.09 $\pm$ 0.06  &        &                   &                &      \\ 
               &                & R     &  0.06 $\pm$ 0.06  &        &                   &                &      \\
               &                & I     &  0.24 $\pm$ 0.07  &        &                   &                &      \\
HD~94851       &  2454950.23292 & V     &                   &        & 0.16 $\pm$ 0.28  &                & \cite{go10a}\\
HD~94851       &  2454952.13280 & V     &                   &        & 0.11 $\pm$ 0.12  &                & \cite{go10a}\\
\hline
HD~65583       & 2455649.09875  & B     & 0.06 $\pm$ 0.08   &        &                   &                &    \\
               &                & V     & 0.05 $\pm$ 0.06   &        &                   &                &     \\ 
               &                & R     & 0.07 $\pm$ 0.24   &        &                   &                &    \\
               &                & I     & 0.09 $\pm$ 0.09   &        &                   &                &    \\
\hline
HD~160529      & 2455649.49125  & B     & 7.46 $\pm$ 0.04   & 20.1   & 7.24              &                & \cite{cl98}\\
               &                & V     & 7.76 $\pm$ 0.07   & 20.4   & 7.52              & 20.1           & \cite{cl98}\\
               &                & R     & 7.41 $\pm$ 0.06   & 21.7   &                   &                &    \\
               &                & I     & 5.49 $\pm$ 0.03   & 21.1   &                   &                &  \\
HD~160529      & 2455650.47868  & B     & 7.36 $\pm$ 0.10   & 19.8    & 6.97              &                & \cite{hs82}\\
               &                & V     & 7.50 $\pm$ 0.12   & 18.7    & 7.31              &                & \cite{hs82}\\
               &                & R     & 7.41 $\pm$ 0.05   & 21.6    &                   &                &     \\
               &                & I     & 4.88 $\pm$ 0.03   & 21.4    &                   &                &   \\
HD~160529      & 2454952.46513  &   V   &                   &                & 7.46 $\pm$0.02    & 20.9    & \cite{go10a}\\
HD~160529      &                &       &                   &                & 7.35 $\pm$ 0.55   &         & \cite{re98}\\ 
HD~160529      &                &       &                   &                & 7.3 $\pm$ 0.54    & 20      & \cite{se75}\\
\hline
\end{tabular}

Our estimates are listed in columns (4) and (5)\\
Estimates  listed  in columns (6) and (7) are  from references listed in column (8).\\
\end{table*}
}

{\footnotesize
\begin{table*}
{\bf Table 3:  $BVRI$ Polarimetry of carbon stars }\\
\begin{tabular}{  c | c | c | c | c |c| c | c  }
\hline
Star names     & E(B-V)  &  Reference  &  HJD & Filter &  p           & $\theta$ &        \\
               &         &             &       &       &  ($\%$)        & (degree)         &    \\
\hline
HE~0310$+$0059 & ----    &    &2455594.16819  & B   & \bf {0.77 $\pm$ 0.10}  & 130.1     &        \\
               &         &    &               & V   & \bf { 0.89 $\pm$ 0.07}  & 130.9      &                \\
               &         &    &               & R   & \bf { 0.70 $\pm$ 0.07}  & 142.1      &                 \\
               &         &    &               & I   & \bf { 0.58 $\pm$ 0.09}  & 147.4      &                 \\
               &         &    &               &     &                     &                &         \\
HE~0518$-$2322 & 0.028   & 1  & 2455595.25775 & B   & \bf {0.39 $\pm$ 0.12}  & 147.5      &       \\
               &         &    &               & V   & \bf {0.33 $\pm$ 0.06}  & 177.9       &               \\
               &         &    &               & R   &  0.33 $\pm$ 0.14  & 144.3      &             \\
               &         &    &               & I   &  0.73 $\pm$ 0.19  &  26.2     &              \\
               &         &    &               &     &                     &                &            \\
HE~0519$-$2053 &  ----   &    & 2455595.21464 & B   &  \bf {0.57 $\pm$ 0.13}  & 159.9      &      \\
               &         &    &               & V   &  0.28 $\pm$ 0.13  & 172.9       &               \\
               &         &    &               & R   &  \bf {0.21 $\pm$ 0.07}  & 147.0      &               \\
               &         &    &               & I   &  0.28 $\pm$ 0.16  & 147.5      &              \\
               &         &    &               &     &                     &                &             \\
HE~0915$-$0327 & 0.041   &  1 & 2455649.31148 & B   &  0.75 $\pm$ 0.32  & 147.5      &         \\
               &         &    &               & V   &  \bf {0.52 $\pm$ 0.07}  & 14.3      &              \\
               &         &    &               & R   &  \bf {0.44 $\pm$ 0.08}  & 147.4      &             \\
               &         &    &               & I   &  \bf {0.77 $\pm$ 0.08}  & 9.4      &              \\
               &         &    &               &     &                     &                &             \\
HE~0916$-$0037 & 0.030   &  1 & 2455594.40375 & B   &  0.46 $\pm$ 0.20  & 12.5      &        \\
               &         &    &               & V   &  0.19 $\pm$ 0.08  &       &              \\
               &         &    &               & R   &  0.25 $\pm$ 0.09  & 127.5      &             \\
               &         &    &               & I   &  0.10 $\pm$ 0.11  &        &              \\
               &         &    &               &     &                     &                &              \\
HE~1030$-$1518 &  0.065  & 1  & 2455650.26381 & B   &  \bf {0.51 $\pm$ 0.06}  & 17.0      &           \\
               &         &    &               & V   &  0.18 $\pm$ 0.09  &       &               \\
               &         &    &               & R   &  \bf {0.49 $\pm$ 0.08}  & 25.3      &              \\
               &         &    &               & I   &  \bf {0.46 $\pm$ 0.07}  & 129.3      &              \\
               &         &    &               &     &                     &                &          \\
HE~1056$-$1855 & 0.045   & 1  & 2455595.35264 & B   &  0.08 $\pm$ 0.08   &       &       \\
               &         &    &               & V   &  0.16 $\pm$ 0.13  &       &               \\
               &         &    &               & R   &  0.19 $\pm$ 0.11  &       &             \\
               &         &    &               & I   &  0.19 $\pm$ 0.11  &       &             \\
               &         &    &               &     &                     &                &          \\
HE~1345$-$2616 & ----    &    & 2455650.39694 & B   & \bf {0.43 $\pm$ 0.05}  &  153.9     &          \\
               &         &    &               & V   & \bf {0.47 $\pm$ 0.06}  &  148.5     &         \\
               &         &    &               & R   & \bf {0.42 $\pm$ 0.07}  &  163.7      &           \\
               &         &    &               & I   & 0.29 $\pm$ 0.10   &  134.2     &           \\
               &         &    &               &     &                     &                &        \\
HE~1428$-$1950 & 0.044   & 1   & 2455594.49870 & B   & 0.34 $\pm$ 0.15   & 179.0       &          \\
               &         &     &               & V   & \bf {0.87 $\pm$ 0.07}   & 150.2      &         \\
               &         &    &               & R   & \bf {0.87 $\pm$ 0.12}  & 152.3     &            \\
               &         &    &               & I   & \bf {0.70 $\pm$ 0.13}   & 150.8      &            \\
               &         &    &               &     &                     &                &           \\
\hline
\end{tabular}

\end{table*}
}

{\footnotesize
\begin{table*}
{\bf Table 3:  BVRI Polarimetry of carbon stars (cont.) }\\
\begin{tabular}{ | c | c | c | c | c |c| c | c   }
\hline
Star names     & E(B-V)  & Reference & HJD    & Filter  & p           & $\theta$ &     \\
               &         &             &       &       &  ($\%$)        & (degree)         &    \\
\hline
HD~1429$-$1411 & ----    &    & 2455649.45900 & B   & \bf {0.69 $\pm$ 0.08}   & 20.0       &         \\
               &         &    &               & V   & \bf {0.90 $\pm$ 0.05}   & 139.6      &           \\
               &         &    &               & R   & 0.67 $\pm$ 0.26   & 126.9      &          \\
               &         &    &               & I   & \bf {0.56 $\pm$ 0.04}  & 129.7      &            \\
               &         &    &               &     &                     &                &        \\
HD~55496       & 0.05    & 2  & 2455649.13837 & B   & \bf {0.32 $\pm$ 0.09}   & 167.9       &         \\
               &         &    &               & V   & \bf {0.18 $\pm$ 0.06}   &       &              \\
               &         &    &               & R   & 0.03 $\pm$ 0.07   &       &             \\
               &         &    &               & I   & 0.16 $\pm$ 0.07   &       &               \\
               &         &    &               &     &                     &                &          \\
HD~81192       &  0.0    & 3  & 2455649.23910 & B   & 0.14 $\pm$ 0.08   &        &       \\
               &         &    &               & V   & 0.17 $\pm$ 0.09   &        &            \\
               &         &    &               & R   & 0.09 $\pm$ 0.06   &        &             \\
               &         &    &               & I   & \bf {0.16 $\pm$ 0.05}   &        &             \\
               &         &    &               &     &                     &                &      \\
HD~92545       &   0.014 & 4  & 2455649.21908 & B   & 0.08 $\pm$ 0.09   &       &          \\
               &         &    &               & V   & 0.02 $\pm$ 0.08   &       &          \\
               &         &    &               & R   & 0.05 $\pm$ 0.08   &       &          \\
               &         &    &               & I   & 0.09 $\pm$ 0.13   &       &           \\
               &         &    &               &     &                     &                &       \\
HD~100764      &  0.02   & 5  & 2455595.37625 & B   & 0.12 $\pm$ 0.08  &           &          \\
               &         &    &               & V   & 0.16 $\pm$ 0.11   &       &             \\
               &         &    &               & R   & 0.28 $\pm$ 0.13   & 137.2      &          \\
               &         &    &               & I   & 0.37 $\pm$ 0.14   &  24.1     &            \\
               &         &    &               &     &                     &                &          \\
LEE 107        & ---     &    & 2455595.40695 & B   & 0.07 $\pm$ 0.09   &         &             \\
               &         &    &               & V   & 0.16 $\pm$ 0.11   &          &                  \\
               &         &    &               & R   & 0.22 $\pm$ 0.13   & 25.7      &             \\
               &         &    &               & I   & 0.31 $\pm$ 0.15   & 16.9      &               \\
               &         &    &               &     &                     &                &            \\
HD~107574      &  0.012  &  4 & 2455649.36574 & B   & 0.05 $\pm$ 0.06   &      &       \\
               &         &    &               & V   & 0.07 $\pm$ 0.07   &          &                \\
               &         &    &               & R   & 0.14 $\pm$ 0.11   &          &              \\
               &         &    &               & I   & 0.19 $\pm$ 0.16   &           &             \\
               &         &    &               &     &                     &                &          \\
HD~111721      &  0.01   &  6  & 2455649.39645 & B   & 0.28 $\pm$ 0.10   &  167.6      &        \\
               &         &    &               & V   & \bf {0.22 $\pm$ 0.04}   &  147.0     &           \\
               &         &    &               & R   & \bf {0.27 $\pm$ 0.07}   &  179.0      &           \\
               &         &    &               & I   & 0.17 $\pm$ 0.07   &         &          \\
               &         &    &               &     &                     &                &        \\
HD~112869      & 0.0     &  5 & 2455650.29685 & B   & 0.13 $\pm$ 0.07   &        &         \\
               &         &    &               & V   & \bf {0.29 $\pm$ 0.08}   & 147.5      &          \\
               &         &    &               & R   & 0.15 $\pm$ 0.12   &          &           \\
               &         &    &               & I   & 0.05 $\pm$ 0.10   &       &          \\
               &         &    &               &     &                     &                &          \\
\hline
\end{tabular}

\end{table*}
}

{\footnotesize
\begin{table*}
{\bf Table 3:  BVRI Polarimetry of carbon stars (cont.) }\\
\begin{tabular}{ | c | c | c | c | c |c| c | c    }
\hline
Star names     & E(B-V)  & Reference & HJD    & Filter  & p           & $\theta$ &    \\
               &         &             &       &       &  ($\%$)        & (degree)         &    \\
\hline
HD~121447      &   0.05  & 7  & 2455650.33307 & B   & 0.29 $\pm$ 0.12   & 147.5      &        \\
               &         &    &               & V   & \bf {0.60 $\pm$ 0.09}  & 166.7       &          \\
               &         &    &               & R   & \bf {0.53 $\pm$ 0.07}   & 166.2       &       \\
               &         &    &               & I   & \bf {0.46 $\pm$ 0.09}   & 165.7       &            \\
               &         &    &               &     &                     &                &         \\
HD~125079      &  0.0    & 8  & 2455650.42737 & B   & 0.28 $\pm$ 0.13   & 152.5      &      \\
               &         &    &               & V   & 0.14 $\pm$ 0.07   &          &           \\
               &         &    &               & R   & 0.37 $\pm$ 0.13   & 139.8      &          \\
               &         &    &               & I   & 0.17 $\pm$ 0.17   &         &             \\
               &         &    &               &     &                     &                &      \\
HD~126681      & 0.0     & 9  & 2455650.45205 & B   & 0.07 $\pm$ 0.07   &        &          \\
               &         &    &               & V   & 0.06 $\pm$ 0.08   &       &         \\
               &         &    &               & R   & 0.18 $\pm$ 0.10   &          &          \\
               &         &    &               & I   & 0.15 $\pm$ 0.14   &       &          \\
               &         &    &               &          &                &                &         \\
HD~164922      &  0.0    & 3  & 2455649.50958 & B   & 0.28 $\pm$ 0.12   & 17.5      &         \\
               &         &    &               & V   & 0.28 $\pm$ 0.10   & 165.3       &           \\
               &         &    &               & R   & 0.15 $\pm$ 0.08   &          &            \\
               &         &    &               &     &                     &                &          \\
\hline
\end{tabular}

{\bf References.} 1. \cite{be07}; 2. \cite{be00}; 3. \cite{ma10}; 4. \cite{a06a};
5. \cite{eg72};  6. \cite{ry95}; 7. \cite{ber99}; 8. \cite{sm93}; 9. \cite{me10}\\
Estimates with p\% ${\ge}$ 3${\sigma}$ are indicated using boldface.\\
\end{table*}
}

\begin{figure}
\resizebox{\hsize}{!}{\includegraphics{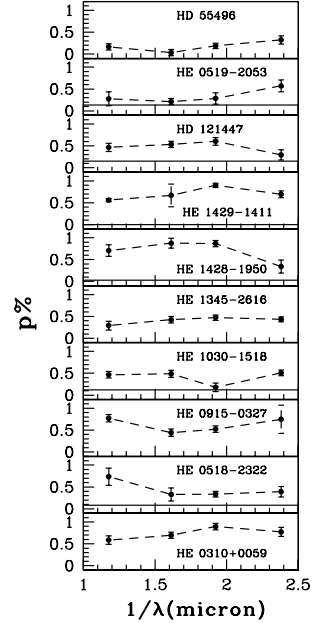}}
\caption{  Response of the observed $BVRI$ percentage polarization 
estimates  with respect to the inverse of the effective wavelength of 
observation  for a sample of CEMP stars and  CH stars.  The 
solid line in each panel indicates the mean interstellar polarization 
level (${\sim}$ 3E(B-V)).  
}
\end{figure}

\section{Results  and discussions}
 Close agreement between our polarimetric estimates for the polarized 
and zero-polarization standard stars with those from the literature lends 
support to the reliability of our results. Estimated p\% for the programme
stars  listed in Table 3 have  not been corrected for interstellar polarization. 
The interstellar reddening listed  for the programme stars is  also low. 
For objects  at low Galactic latitude, intrinsically polarized stars may 
have a component of interstellar polarization  that  is negligible for 
objects at high Galactic latitude. Estimated errors in p\%  are 
photon-noise-dominated and  negligibly small  in the case of very bright 
objects.

  Taking the error estimates into account,  the
derived percentage polarization  is above the 3${\sigma}$ limit in all
four bands for only one object, HE~0310+0059. Seven stars exhibit 
polarization  above the 3${\sigma}$ limit in three bands, three stars 
in two bands, and two stars  in a single band. Nine objects in our sample 
show polarization at a level  below the 3${\sigma}$ limit in all four bands. 
The polarization estimates above the 3${\sigma}$ limit are indicated using 
boldface in Table 3.
\subsection {Polarimetric characteristics: wavelength dependence of 
polarization}

The polarization estimates for the programme  stars are low ($\le$ 1\%)  
(Table 3).  In figure 1, we  show  the response of the $BVRI$ 
polarimetric  estimates for a few selected stars that show polarization
above the 3${\sigma}$ limit in more than two bands  against the corresponding  
inverse of the effective wavelength of observation.  The wavelength 
dependence of polarization is  found to be random in nature and does  not 
follow any systematic pattern. 
  A discussion of  the wavelength dependence of polarization is 
meaningful  only in cases where the estimates of  percentage polarization are
above the 3${\sigma}$ limit. Nine objects in our sample,
 HE~0916$-$0037, HE~1056$-$1855, HD~92545,
HD~100764, LEE~107, HD~107574, HD~125079, HD~126681, and HD~164922, 
show polarization below  the 3${\sigma}$ limit in all four bands.
  The object, HE~0310+0059, 
that shows significant polarization in all  four  bands shows a 
maximum  in the V-band. 
The objects HD~81192 and HD~112869  show
polarization above  the 3${\sigma}$ limit  only in  one band, the 
I band and V band, respectively.
 HD~55496, HD~111721, and HE 0519$-$2053 show polarization 
above  the 3${\sigma}$ limit in  two bands, BV, VR, and BR, respectively.
Seven  objects in our sample, HE~0518$-$2322, HE~0915$-$0327, 
HE~1030$-$1518, HE~1345$-$2616,
HE~1428$-$1950, HE~1429$-$1411, and HD~121447 show  polarization
above  the 3${\sigma}$ limit  in three bands. While the objects
HE~1030$-$1518, HE~1345$-$2616, and HE~1428$-$1950 show a rather flat
wavelength dependence of polarization within the errors in B, R, and I bands,
the objects  HE~0518$-$2322 and  HE~0915$-$0327 show a 
maximum in I band, and the objects HE~1429$-$1411, and HD~121447
show a maximum in V band. 

 As discussed in Zickgraph \& Schulte-Ladbeck (1989), dust particles of 
different radii and composition can show polarization maximum in  different 
bands. In many cases maximum polarization for scattering of light caused 
by small dust particles can be observed at short wavelengths. Small carbon 
particles are known to exist in the circumstellar envelopes of carbon stars.
The optical polarization of the evolved carbon star R Scl showed 
a ${\sim}$ ${\lambda}^{-4}$ wavelength-dependence, attributed to
scattering by small amorphous carbon dust grains (Yudin \& Evans (2002)) and
polarimetric variability was also detected  for this object 
on time-scales from hours to years.

 In the  upper  panel of figure 2,  we  show  the V-band percentage
polarization estimates  (p$_{v}$\%) of the same stars as
shown in figure 1,  against the
interstellar reddening E(B-V) values whenever available. The mean 
interstellar polarization  ${\sim}$ 3E(B-V) (\cite{pa05}) is also 
indicated.  The estimated  V-band percentage polarization is higher
than the mean interstellar contribution to the polarization, 
indicating that they are intrinsically polarized. 

 Polarization of carbon stars is  known to show  flatter wavelength 
dependence than the oxygen-rich stars. A mild wavelength dependence 
of polarization noticed in some cases  is  somewhat similar to those 
reported by 
Raveendran (1991) for  a few carbon stars of Mira variables; i.e., 
RT Pup shows a systematic increase in polarization towards the red, 
and X Vel shows higher  polarization in the R band than in the 
V band. Polarimetric studies of some giants and supergiants 
(Dyck \& Jennings (1971), Lopez \& Hiriart (2011)) have shown that the
wavelength dependence of polarization varies widely and often
changes with time and that the largest changes in the amount of
polarization occurs in the ultraviolet. There are,  however,  exceptional
cases of  S~Per and 119 Tau. In S~Per  polarization rises abruptly
into the ultraviolet but at other times decreases on  a timescale
of few days (Dyck \& Jennings (1971)). In the case of 119~Tau, polarization 
in the U-band was found to be about one-half that of the V-band;  in addition,
the position-angle was  found to show changes with wavelength, 
 which they  attribute to  the presence of more 
than one component of polarization. Cases where changes in the position 
angle with wavelength range from almost nothing to as high as 90 degrees  
are also noted  by Dyck \& Jennings (1971). Monitoring time variations 
of position angles with respect to wavelength would be worthwhile for
investigating  whether  there is  a tight  relationship between these two 
parameters.

\subsection {Polarimetric characteristics: temperature  dependence of 
polarization}

 The response of the V-band polarimetric estimates (p$_{v}$\%) for the 
programme  stars with respect to their effective temperatures is shown in 
figure 2 (lower panel). Observed p$_{v}$\% estimates do not seem to 
have any correlation with temperature.

Apart from two objects,  HE~ 0915$-$0327 and HE~1429$-$1411 with 
effective temperatures of 2920 K and 3058 K, respectively, the  
temperature for the rest of  the CEMP stars ranges from 4200 to 
4900 K. The coolest CEMP star in our sample,  HE~ 0915$-$0327,  with 
p$_v$ $\sim$ 0.5$\% $ shows a maximum polarization in the I-band (0.77\%). 
Within the sample the  highest polarization is seen in HE~1429$-$1411 
with p$_v$ $\sim$ 0.9$\% $; it shows  minimum polarization in the  I-band. 
The temperature and metallicity ranges for the HD stars  are about 
3200 - 6400 K, and 0.05 to $-$1.6. The hottest objects HD~92545 
(6240 K) and HD~107574 (6340 K) do not show any significant 
polarization in BVRI. The coolest object, HD~112869, shows 
polarization  only in the V-band with p$_v$ ${\sim}$ 0.29$\% $. HD~121447
with  T$_{eff}$ of about 4200 K returns the highest polarization
in the V-band among all the HD stars in our sample.

The metallicity data for this sample of stars is scanty; the metallicity 
estimates are available only for  three CEMP stars and nine HD stars 
in our sample (Table 4), and   a fraction of them do not show any 
significant polarization.  To examine the temperature and 
metallicity dependence of polarization,  it would be necessary to 
consider a larger sample covering a much wider range in temperature 
and metallicity.  Among the objects that show finite polarization,  both  
CH  and CEMP  stars show similar responses  with respect to T$_{eff}$.

 Simultaneous polarimetric observations in BVRI are limited in the 
literature. Except for HD~100764 for none of the programme  stars'   
literature values are available. We have presented first-time 
polarimetric estimates for this set of  objects,  which  would be 
 useful later for time variability studies of  polarization and  its
wavelength dependence,  as well as for the  time variability of polarization 
and position angles.

\begin{figure}
\resizebox{\hsize}{!}{\includegraphics{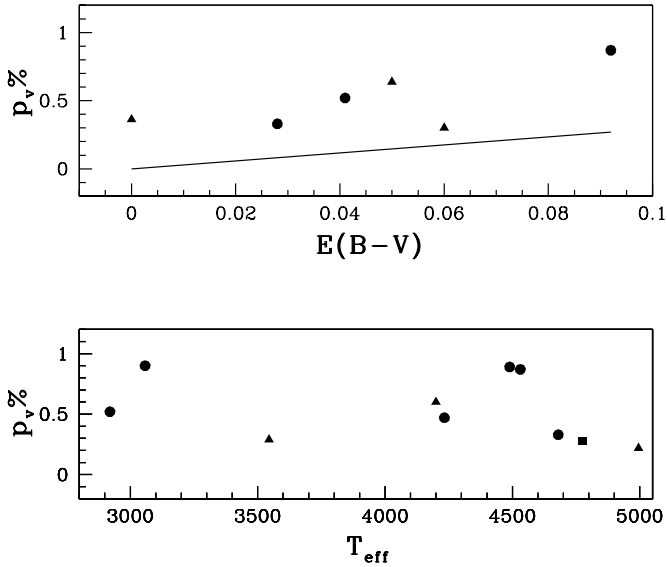}}
\caption{
Top: the observed p$_{v}$\%  vs  the
interstellar E(B-V). The solid line represents the mean interstellar 
polarization (${\sim}$ 3E(B-V)). E(B-V) values  are not 
available  for all  ten  objects. In all the panels 
the stars from the CH star catalogue are represented by solid triangles 
and the carbon stars from Hamburg/ESO survey  by solid circles. The
star represented by a solid square corresponds to HE~0519-2053.
 Bottom: the observed  V-band percentage polarization (p$_{v}$\%)
  vs the  effective temperatures 
for the same objects as in figure 1. Polarization does not seem to 
have any correlation with  temperature. 
}

\end{figure}

\noindent
\subsection {Comments on individual stars}
{\it {F Str ${\lambda}$ 4077 stars HD~92545 and HD~107574}}\\
 These two  objects  are known to  show abnormally strong 
Sr line at 4077\AA\,  in their spectra. Bidelmann (1981) classified these  
objects  as F Str ${\lambda}$ 4077 stars. They exhibit   enhancement of 
light and heavy s-process elements but  abundances of iron-peak elements 
are similar to those generally seen in F type stars. The radial velocities  
of HD~92545 and HD~107574 are  $-18$ km s$^{-1}$ and  
$-28$ km s$^{-1}$, respectively.
 HD~107574 is said  to be a binary (\cite{no91}). The estimated 
distance for HD~92545 is about 101 pc and for HD~107574 the distance 
ranges between  120 - 150 pc (North (1987)). Based on kinematics and 
chemical compositions,  \cite{no91} suggest that  these 
two objects are probably the main-sequence progenitors of barium stars  
rather than being  population II objects. Estimated $BVRI$ percentage 
polarization for these two objects are below the 3${\sigma}$ limit in 
all  four bands.
The near-zero polarization observed in these two objects may be an  
indication of  their early stage of evolution, yet to form  aspherical 
dust shells around them. At the instance of having circumstellar 
envelopes this would imply they are  spherically symmetric.

\noindent
{\it The subgiant CH stars HD~55496 and  HD~125079}\\
HD~55496 and  HD~125079 are entered  in both the barium 
star catalogue of \cite{lu91}  and the CH star catalogue of \cite{ba96} 
indicating  that  some uncertainties exist  regarding  their 
classification. These two objects are referred to as subgiant CH stars by 
Smith \& Lambert (1986). A high radial velocity of $\sim$ $+$324 km s$^{-1}$ 
reported by Bond (1974) for HD~55496  is consistent with those generally 
observed for CH giants. In view of the close similarities between
the spectral peculiarities of the two groups,  it was  proposed that 
the subgiant and giant CH stars form a continuous sequence of stars of 
the same abnormal chemical composition, differing only in temperature 
and  luminosities  (Bond (1974)).

\noindent
{\it HD 100764}\\
This object is known to have  a detached cold dust shell that  is 
optically thin with a  relatively large dust grain size (\cite{pa91}). 
HD~100764 is the only example in our sample for which previous 
polarimetric estimates exist. Multiband photo-polarimetry does  not show
any  significant variations in  polarization  in B, V, R,  and I bands. 
A comparison of the polarimetric estimates for this object is presented in 
Table 5.

\noindent
{\it   HD~121447, HD~111721} \\
The  object HD~121447  is reported  to exhibit an enhancement of 
nitrogen abundance (roughly twice solar), slight depletion of oxygen,  
and near solar or marginal enhancement of carbon abundances (Smith (1984)). 
Abundance peculiarities, such as excess $^{12}$C and s-process
overabundances observed in this star,  are accounted for by  assuming that
roughly 10$^{-2}$ of the outer envelope is contaminated by pure $^{12}$C
and s-process heavy elements (\cite{sm84}). E(B-V) for this object is
small $\sim$ 0.05; the estimated polarizations  seem to represent the
star's intrinsic polarization property.

The object HD~111721 has been referred to as a metal-poor subgiant star by
\cite{ca97}. However, from kinematics,  \cite{ry95}  confirm
it to be a  Galactic halo object.
\cite{gr91}  determine a radial velocity of $+$22.3 km s$^{-1}$ 
for this object. This object shows polarization above the 3${\sigma}$ limit
in the V and R bands.

\noindent
{\it HE~0310+0059, HE~0518$-$2322,  HE~0519$-$2053, 
HE~0915$-$0327,  HE~1030$-$1518, HE~1345$-$2616,
 HE~1428$-$1950, HE~1429-1411}\\ 
Medium-resolution spectral analyses and atmospheric parameters for a 
number of these objects can be found   in the literature 
(\cite{go05}, Goswami et al. (2007), (2010b), \cite{ken11}). The spectra of 
HE~0310+0059, HE~0518$-$2322, and  HE~1056$-$1855 closely resemble the 
spectrum of HD~26, a well-known classical CH star (\cite{go10b}). The 
atmospheric parameters  T$_{eff}$, log\,{g}, and  metallicity [Fe/H] for  
HE~0310+0059 are, respectively, 4861 K, 1.69,  and $-$1.32. This object 
exhibits  polarization above the 3${\sigma}$ limit in all  four bands. The 
atmospheric parameters for HE~0519$-$2053 are very similar (Table 4) to  
those of HE~0310+0059; however, HE~0519$-$2053 shows polarization above  
the 3${\sigma}$ limit only in the B and  R bands. HE~0519$-$2053 and 
HE~0310+0059 both exhibit enhancement of carbon with [C/Fe] values  
0.95 and 1.24, respectively (\cite{ken11}). The object HE~1428$-$1950 
is also enhanced in carbon with [C/Fe] = 1.78. The estimated p\% in  
$BVRI$ for HE~0310+0059 is  very similar to the estimates obtained  
for  HE~1429$-$1411. The polarimetric estimates for HE~0518$-$2322, 
HE~0519$-$2053, HE~0915$-$0327, HE~0916$-$0037, HE~1030$-$1518,  and 
HE~1345$-$2616 are also within a range of 0.2\% to 0.9\%. The object 
HE~1428$-$1950  is classified as a C-R star  (\cite{go10b}).  The 
estimates of p\% in $BVRI$ for this object are not very  different from 
those of other HE stars and are within the above range.

\begin{table*}
{\bf Table 4: Atmospheric parameters of the programme  stars }\\
\begin{tabular}{  c| c| c| c| c    }
\hline
Star names     & T$_{eff}$ & log$g$       & [Fe/H] & Reference     \\
               &    (K)      & (cm s$^{-2}$)  &  (dex)     &               \\
\hline
HE 0310+0059      &  4489$^a$/4861    &  1.69    & $-$1.32      & 1   \\
                  &      &      &      &          \\
HE 0518-2322      & 4680$^a$     &  --- & ---  & --- \\ 
                  &      &      &      &          \\
HE 0519-2053      & 4775     &  1.46    & $-$1.45     &  1  \\ 
                  &      &      &      &          \\
HE 0915-0327      & 2920$^a$     & --- & ---  &  --- \\
                  &      &      &      &          \\
HE 0916-0037      & 4872$^a$     & ---  & ---   &  --- \\ 
                  &      &      &      &          \\
HE 1030-1518      &  4350$^a$     & ---  & ---  & ---    \\
                  &      &      &      &          \\
HE 1056-1855      & 4280$^a$     & ---   & ---  & ---   \\
                  &      &      &      &          \\
HE 1345-2616      &  4233$^a$     & ---     & ---  &  ---      \\
                  &      &      &      &          \\
HE 1428-1950      & 4531     & 0.85     & $-$2.07      &  1    \\
                  &      &      &      &          \\
HE 1429-1411      &  3058$^a$    & ---    & ---     & ---   \\
                  &      &      &      &          \\
HD 55496          & 4800 & 2.8 & -1.55 &  2 \\
                  &      &      &      &      \\
HD 81192          & 4705 & 2.50& -0.62 & 2   \\
                  & 4582 & 2.75& -0.70 & 3   \\ 
                  & 4755 & 2.4 & -0.60 & 4  \\
                  & 4755 & 2.4 & -0.61 & 5  \\
                  &      &      &      &    \\
HD 92545          & 6240 &  4.23 &-0.26 & 6     \\
                  &      &      &      &           \\
HD 100764         & 4846 &  2.2 & -0.59 & 7   \\
                  &      &      &      &          \\
LEE 107           & 3288$^a$     & ---  & ---  &  ---    \\
                  &      &      &      &             \\
HD 107574         & 6340 & 3.87 &-0.36 & 6  \\
                  & 6340 & 3.87 &-0.54 & 6        \\
                  & 6340 & 3.87 &-0.16 & 6         \\
                  &      &      &      &            \\
HD 111721         & 4995 & 2.52& -1.26 & 8   \\
                  & 4825 & 2.2 & -1.54&  9        \\
                  & 4800 & 3.00& -1.68 & 10         \\
                  & 5164 & 3.27& -0.98 & 11         \\
                  & 4940 & 2.40& -1.34 & 12          \\
                  & 5103 & 2.87& -1.25 & 13        \\
                  & 5000 &     & -1.34 & 14         \\
                  & 4860 & 2.2 & -1.57 & 15        \\
                  &      &      &      &            \\
\hline
\end{tabular}
\end{table*}

{\footnotesize
\begin{table*}
{\bf Table 4: Continued }\\
\begin{tabular}{  c| c| c| c| c    }
\hline
Star names     & T$_{eff}$ & log$g$       & [Fe/H] & Reference     \\
               &    (K)      & (cm s$^{-2}$)  &  (dex)     &               \\
\hline
HD 112869        &  3545$^a$          &  ---      &  ---      &   ---      \\  
                 &      &      &      &          \\
HD 121447        & 4200 & 0.8 &  0.05 & 16       \\
                 &      &      &      &          \\
HD 125079        & 5300 & 3.50 &-0.16&  17    \\
                 & 5305 &  3.50 &-0.30  &  18    \\
                 &      &      &      &          \\
HD 126681        &  5450 & 4.5 & -1.25  &9     \\
                 &  5533 & 4.2 & -1.14  &19           \\
                 &  5595 & 4.43& -1.12 & 20        \\
                 &  5625 & 4.95 &-1.09 & 11          \\
                 &  5500 & 4.63 &-1.45 & 21      \\
                 &      &      &      &          \\
HD 164922        & 5412$^a$    &  ---  &  ---  &  ---      \\    
\hline
\end{tabular}

{\bf References.} 1: \cite{ken11}; 2. \cite{ce07}; 
3. \cite{co86}; 4. \cite{luck85}; 5. \cite{luck83}; 
6.  \cite{no94}; 7.  \cite{do84}; 8.\cite{gr94};
9.\cite{fu00}; 10. \cite{ca97}; 11. \cite{gr96}; 
12.\cite{ry95}; 13. \cite{gr94}; 14. \cite{pi93};
15. \cite{fr88}; 16. \cite{sm84}; 17.  \cite{sm93}; 18. \cite{sm86};
19. \cite{ni00}; 20. \cite{ni97}; 21. \cite{to92};\\
 $^a$  estimated from (J-K) 
temperature calibrations of \cite{al96}.

\end{table*}
}

{\footnotesize
\begin{table*}
{\bf Table 5:  HD~100764: comparison of  polarimetric estimates  with literature values}\\
\begin{tabular}{  c | c | c | c | c | c| c| c | c  c | c  }
\hline
Star name     & E(B-V)  & HJD        & Filter   & p           & $\theta$ & HJD       & Filter & p     & $\theta$  & Reference \\
              &         &            &          & ($\%$)         &  (degree)   &        &       & ($\%$)   & (degree)  &  \\
(1)            &  (2)   &  (3)     & (4)             &  (5)           &  (6)        &   (7)  & (8)      &   (9)        & (10)  &   \\
\hline
               &             &          &                 &                &             &         &         &                 &  &       \\
HD~100764      &0.02 &  2,447,971    & B      &  0.13$\pm$ 0.12& 127.2 $\pm$ 25.1 & 2455595.37625& B &  0.115 $\pm$ 0.0860 & 143.4 & 1  \\
            &     &2,447,971    & V      &  0.20$\pm$ 0.05&              &   2455595.37625 & V  & 0.163 $\pm$ 0.107  & 143.4 &  1 \\
            &     &2,447,971    & R      &  0.01$\pm$ 0.04& 102.6 $\pm$ 6.0  & 2455595.37625& R &  0.278 $\pm$ 0.132  & 137.2 & 1  \\
            &     &2,447,971    & I      &  0.40$\pm$ 0.08& 78.0 $\pm$ 5.5   & 2455595.37625& I &  0.372 $\pm$ 0.144  & 24.1 & 1   \\
            &     &2,448,398    & V      &  0.42$\pm$ 0.14&                  & 2454952.14664& V & 0.330 $\pm$ 0.06    &   4.0   & 2 \\
               &     &         &        &                &                  &              &   &                     &           &    \\
\hline
\end{tabular}

Estimates  listed in columns (5) and (6) are from \cite{pa05}\\
Our estimates corresponding to the HJDs in column 7 are presented in 
columns 9 and 10\\
In the last column, reference 1 indicates this work and 2 
indicates \cite{go10a} \\
\end{table*}
}

\section{Conclusions}
CH and CEMP-s  stars have been claimed to share similar elemental abundance
properties,  and  this implies any possible effects of varying chemical 
composition  on the polarization data are  minimal.  
We  used first-time $BVRI$ polarimetric observations
for the sample of CH and CEMP stars to examine whether  they also share 
 polarimetric similarities. 
 The  polarimetric estimates  of the objects 
are low with a range of $\sim$   0.1 to 0.9 \% and  do not show any 
systematic dependence on the inverse of the  effective  wavelength of 
observation.
 In the plane of 
differential polarization,   defined as the difference between
the maximum and the minimum polarizations within the $BVRI$-bands, versus
their visual magnitude,  the CH and CEMP stars 
occupy a  narrow band. 
 The standard deviation for  the distribution of the differential 
polarization is $\sim$  0.036 for the objects shown in figure 1.
 Polarization properties of the stars can be used to infer
the evolutionary stage,  as well as the presence of circumstellar envelopes;
expressing this in terms of their differential polarization could be useful 
for  understanding the measure of inhomogeneity of the circumstellar dust 
composition. For instance, it would be interesting to examine 
 two extreme cases of a high and 
a low value in differential polarization, if the high value  indicated
 greater inhomogeneity than  the lower value.  It 
would be worthwhile examining the time dependence of  this behaviour 
by considering a  larger sample.

 The objects HD~55496, HD~125079, and  HD~121447 have multiple  identities 
in the literature,  as listed  in the CH star catalogue
(Bartkevicious (1996)),  as well as in the barium star catalogue (L\"u (1991)). 
Chemical composition studies  have shown that   Ba stars have 
same s-process signature as AGB stars and that they exhibit [Ba/Fe] and
[Eu/Fe] ratios identical to those of CEMP-s stars. It was therefore suggested 
that  the CEMP-s stars 
and barium stars belong to the same category of AGB
mass-transfer and differ only in  metallicity
(\cite{a06b}, \cite{ma10}). 
 Our polarimetric estimates for these objects 
 are  not too different from the estimates of the other members of the sample.

 Because both  CH stars and  the known members of CEMP-s are 
components of binary systems, it is  reasonable to expect consequent 
asymmetries and inhomogeneities in the circumstellar envelopes,  hence 
 net polarization. 
Stars with circumstellar materials  exhibit a certain amount of linear
polarization. 
Rayleigh scattering in the stellar envelope (Harrington (1969)) and dust 
scattering in the surrounding shells are some  suggested possible mechanisms 
of  polarization. 
 Polarization  increases with asymmetries present in 
the geometry of the shells.  In the case of spherical symmetry, all
polarization vectors cancel, and 
a net zero polarization is observed if the envelope is unresolved. 
The question of dust formation
and the development of a non-spherical structure  in stellar envelopes 
can also be addressed through polarization studies. For gaining insight 
into the distribution of dust,  the geometry and structure of the shells,  
mass-ejection mechanism, and mass-loss rates etc.,  it is required to 
conduct model-based studies of  the observed polarization.
Grain formation  could also lead to  temperature variation  over the 
surface,  causing  asymmetries. In the case of cool stars,  grains can 
condense so close to the stars that internal illunination of a dust shell
by a photosphere with  non-uniform surface brightness  can give rise to
polarization changes across a spectral band (\cite{rav91}); however, 
the amount of changes that can be  produced by this mechanism remains 
to be determined.

A few objects in our sample show polarization
at a level below 0.2\%. It would be interesting to explore how the accreted 
mass gets integrated into the stellar system,  giving rise to a circumstellar
envelope attaining perhaps spherical symmetry  and thus showing null 
polarization.
 Models for  mass loss with dust condensation  (e.g. \cite{wi00}, \cite{ho03})  
assume a spherical geometry with grains
forming in a shell whose radius is determined by temperature, density, 
and condensation fraction with distance from the star. Such models
confirm that during dust condensation the newly formed grains absorb 
stellar photons and are driven outward by radiation pressure. 
While  colliding with the molecules of the cool stellar atmospheres 
the grains exert an outward momentum to the gas,  resulting in a dust-driven 
stellar wind. Observations of both the dust and gas in stellar winds 
show, however, that the distribution of matter is not spherical in 
many cases. Mapping of circumstellar envelopes in a large number of 
cases  using maser line emission of OH  also shows that even the most 
regular envelopes deviate slightly from spherical symmetry,  with the 
wind outflow speed increasing towards the poles (\cite{bo91}).
Multi-epoch polarimetric observations  would be useful to infer the
extent of time-dependent variations in polarization both in the amount 
and in wavelength-dependence.\\

{\it Acknowledgement}\\
 We  thank  the anonymous referee for the  useful suggestions.
This work made use of
the SIMBAD astronomical database, operated at the CDS, Strasbourg, France,
and the NASA ADS, USA. DK is a JRF in the  DST project SR/S2/HEP-09/2007;
funding from this project is gratefully acknowledged. 
\\

\end{document}